\documentstyle[11pt,aaspp4]{article}
\newcommand{\cd}{cm$^{-2}$ }
\newcommand{\phion}{$\Phi _{ion}$}
\newcommand{\kms}{km\, s$^{-1}$}
\newcommand{\e}{$\times$10}
\newcommand{\g}{NGC~3067}
\newcommand{\hsc}{$h_{75}$}
\newcommand{\fesc}{$f_{esc}$}
\newcommand{\ha}{H$\alpha$}
\newcommand{\units}{photons cm$^{-2}$ s$^{-1}$}
\newcommand{\ffe}{$f_{Fe}$}
\newcommand{\fmg}{$f_{Mg}$}

\begin{document}

\title{NEW HST OBSERVATIONS OF THE HALO GAS OF NGC~3067 -- 
LIMITS ON THE EXTRAGALACTIC IONIZING BACKGROUND AT LOW REDSHIFT
AND THE LYMAN CONTINUUM ESCAPE FRACTION}    
\author{JASON TUMLINSON, MARK L. GIROUX, J. MICHAEL SHULL\altaffilmark{1}, \& JOHN T. STOCKE}      
\affil{Center for Astrophysics and Space Astronomy, \\ 
Department of Astrophysical and Planetary Sciences, \\
 University of Colorado, CB
389, Boulder, CO, 80309 \\
Electronic Mail: (tumlinso, giroux, mshull, stocke)@casa.colorado.edu}
\altaffiltext{1}{Also at JILA, University of Colorado and 
National Institute of Standards and Technology.}
\centerline{\today} 
\begin{abstract}
We present ultraviolet spectroscopy from HST/GHRS and reanalyze
existing \ha\ images of the quasar/galaxy pair 3C~232/NGC~3067 and of
the halo gas associated with \g.  The spectra permit measurement of, or
limits on, the column densities of \ion{Fe}{1}, \ion{Fe}{2},
\ion{Mg}{1}, and \ion{Mg}{2} in the absorbing cloud.  Two distinct
models of the extragalactic radiation field are considered:  (1) the
ionizing spectrum is dominated by a power-law extragalactic continuum,
and (2) the power-law spectrum contains a Lyman break, implying
enhanced flux longward of 912 \AA\ relative to the hydrogen-ionizing
flux.  Models of the second type are required to fully explore the
ionization balance of the Fe and Mg in the model cloud.  The
\ha\ images constrain the escape fraction of Lyman continuum photons
from the galaxy to \fesc\ $\leq$ 0.02. With the assumption that the
cloud is shielded from all galactic contributions, we can constrain the
intensity and shape of the extragalactic continuum.  For an
AGN-dominated power-law extragalactic spectrum, we derive a limit on
the extragalactic ionizing flux \phion\ $\geq$ 2600 photons cm$^{-2}$
s$^{-1}$, or $I_{0} \geq 10^{-23}$ erg cm$^{-2}$ s$^{-1}$ Hz$^{-1}$
sr$^{-1}$ for a $\nu ^{-1.8}$ ionizing spectrum and a cloud of constant
density.  When combined with previous upper limits from the absence of
\ha\ recombination emission from intergalactic clouds, our observations
require 2600~$\leq$~\phion\ $\leq$~10000 photons cm$^{-2}$ s$^{-1}$.
We show that if galactic contributions to the incident radiation are
important, it is difficult to constrain \phion.  These results
demonstrate that galactic halo opacities and their wavelength
dependence are crucial to understanding the abundance of low-ionization
metals in the IGM.  
\end{abstract}

\keywords{galaxies: halos --- galaxies: individual (NGC 3067) --- quasars:
individual (3C~232) --- intergalactic medium --- quasars: absorption
lines --- diffuse radiation}

\section{INTRODUCTION}

The extragalactic radiation field and its evolution with redshift play
an important role in our understanding of the intergalactic medium
(IGM) and its constituents. Ionizing photons, believed to be produced
primarily by active galactic nuclei (AGN) but with a potentially large
contribution from starburst galaxies, permeate the IGM and dominate the
ionization of Ly$\alpha$ clouds, QSO metal-line absorbers, and possibly
high-latitude Galactic clouds. Tight constraints on the intensity and
spectrum of this radiation are essential to the interpretation of IGM
observations.

A direct measurement of the intensity of the ionizing background is
impossible due to its attenuation by Galactic \ion{H}{1} and dust.
Thus, we must measure its effects on detectable gas and deduce its
intensity and spectrum from the photon ``fossil record''.  Typically,
conclusions are hindered by the necessary introduction of model
assumptions about the affected gas and by the inherent uncertainties of
photoionization modeling.

Careful treatments of the radiative transfer of ionizing photons within
the IGM (Haardt \& Madau 1996; Fardal et al. 1998; Shull et
al.~1999) have shown that the diminished opacity of the IGM at low
redshift implies that the shape of the extragalactic spectrum at $z
\longrightarrow 0$ approximates the intrinsic spectra of the ionizing
sources.  In contrast, the high-$z$ ionizing background spectrum is
dominated by the effects of radiative transfer in the IGM.  Thus, at
low redshift, a first-cut picture of the extragalactic ionizing
spectrum is the intrinsic spectrum of QSOs, unaffected by attenuation
due to the IGM.  Zheng et al. (1997) combined HST observations to form
a composite spectrum for a radio-quiet QSO which may be fitted by a
power law, $\nu ^{-\alpha _s}$, with spectral index $\alpha _{s} = 1.8
\pm 0.3$ between 350 \AA\ and 1050 \AA.  These results motivate an
initial assumption that the ionizing background has a similar power-law
form.  The relative contribution of ionizing photons produced in OB
associations in galaxies, however, is potentially very large.  The
dimensionless parameter \fesc\ expresses the total fraction of Lyman
continuum (LyC) photons produced in a galaxy that escape its
interstellar medium and halo and contribute to the extragalactic
ionizing background.  If the average \fesc\ $\sim 0.05$ in starburst
galaxies, then their contribution may exceed that of QSOs at low
redshift (Madau \& Shull 1996; Giallongo et al.~1997; Shull et
al.~1999).

To facilitate comparisons between results obtained with different
assumed spectral shapes, we define a quantity $\Phi _{ion}$, the
one-sided, normally incident hydrogen-ionizing photon flux:
\begin{equation} \Phi _{ion} = 2\pi \int_{0}^{1} \mu d \mu
\int_{\nu _{0}}^{\infty}
 I_{\nu} \frac{d \nu}{h \nu} = 2630\, I_{-23} \left(
\frac{1.8}{\alpha _{s}} \right) \rm cm^{-2} \, s^{-1},
\end{equation} where $I_{\nu} =
I_{0}(\nu / \nu_{0} )^{-\alpha _{s}}$ and $I_{0}$ = (10$^{-23}$ erg
cm$^{-2}$ s$^{-1}$ Hz$^{-1}$ sr$^{-1}$)$I_{-23}$ at 1 Ryd.  
In this paper, we will quote \phion\ and assume $\alpha _{s} = 1.8$ for
citations of $I_{-23}$. 

Previous measurements of \phion\ and $I_{-23}$ relied on a number of
different techniques.  Maloney (1993) and Dove \& Shull (1994) analyzed
the \ion{H}{1} edges of galactic disks to infer $10^{4.0}
\leq$~\phion\ $\leq 10^{4.7}$ \units.  Kulkarni \& Fall (1993) used the
falloff in the distribution of Ly$\alpha$ forest clouds close to
quasars (the proximity effect) at $\langle z \rangle \sim 0.5$ to
derive \phion\ $\sim$ 10$^{2.7}$ -- $10^{4.0}$ \units.  In a previous
study of the halo gas associated with NGC~3067, Stocke et al. (1991)
used non-detection of H$\alpha$ emission from the cloud to argue for an
upper limit on \phion\ $\leq 10^{4.9}$ \units.  More recent efforts to
detect H$\alpha$ emission from intergalactic clouds have set upper
limits on \phion\ $\leq$ 10$^{4.0}$ cm$^{-2}$ s$^{-1}$ (Donahue,
Aldering, \& Stocke 1995; Bland-Hawthorn \& Maloney 1999).

The quantity \fesc\ is poorly known.  Leitherer et al.~(1995) used FUV
spectra from the {\em Hopkins Ultraviolet Telescope} to place upper
limits on \fesc\ $\leq$ 0.0095, 0.017, 0.048, and 0.15 for four nearby
starbursts.  Subsequent reanalysis of the same data by Hurwitz et
al.~(1997) argued for higher fractions, \fesc\ $\leq$ 0.032, 0.052,
0.11, and 0.57.  Based on an opacity-dominated theoretical model of
\fesc, Deharveng et al. (1997) argued for a very low \fesc\ $\leq
10^{-4} - 10^{-3}$.  More recently, Bland-Hawthorn \& Maloney (1999)
used measurements of H$\alpha$ emission from the Magellanic Stream to
infer \fesc\ $\sim 0.06$ for our Galaxy.  The paucity of measurements
in the literature can be attributed to the difficulty of such
endeavors;  \fesc\ directly affects the spectra of galaxies in a
wavelength regime obscured by our own ISM, especially for available
targets at low redshift.

When comparing theoretical and observational estimates of \fesc, it is
important to distinguish between the two extant physical conceptions of
the escape fraction. The ``opacity'' interpretation (Deharveng et al.
1997) postulates that galaxies are shrouded in thick layers of
\ion{H}{1} that provide a high photoelectric opacity to ionizing
photons.  In this model, the galactic ionizing photon flux is
attenuated uniformly by optical depth $\tau$, where in the simplest
model $\tau = N_{HI} \sigma _{H}(\nu)$, $N_{HI}$ is the column density
of neutral hydrogen in the absorbing gas, and $\sigma _{H} (\nu)$ is
the photoionization cross-section for neutral hydrogen. The
``geometric'' interpretation (Dove \& Shull 1994; Dove et al.~1999)
rests on essentially the same model of the galactic disk and halo, but
with the addition of ``holes'' produced by supernovae and winds from
massive stars. These holes contain little \ion{H}{1}, and LyC photons
may escape relatively freely through them.  The important distinction
is that, in the first case, \fesc\ is dominated by \ion{H}{1} opacity
and may be uniform for all galaxies of a given morphology.  In the
second case, \fesc\ is dominated by inhomogeneities in the disk and
halo and may be sensitive to an individual galaxy's star formation
history.

With the goal of improving limits on \phion\ and gauging the importance
of radiation longward of 912 \AA, we observed the quasar/galaxy pair
3C~232/NGC~3067 and conducted a study of metal absorption lines in
the halo gas of NGC~3067 using the Goddard High Resolution Spectrograph
(GHRS) on the {\em Hubble Space Telescope}. We measured column
densities of \ion{Fe}{2}, \ion{Mg}{1} and \ion{Mg}{2} and placed limits
on \ion{Fe}{1} in the NGC~3067 halo gas. We interpreted these data to
estimate the extragalactic photoionizing radiation field and to model
the thermal equilibrium of the halo gas.  The technique of using
absorption-line ratios to infer the ionizing background (Stocke et al.
1991) is sensitive to smaller portions of gas in the QSO sightline. The
H$\alpha$ method derives limits on \phion\ from emission measurements
of the entire cloud and is insensitive to its shape and size. While the
non-detections of H$\alpha$ give upper limits on \phion, in principle
our measurements of Fe and Mg yield lower limits.  This method has the
drawback of an additional dependence of the Fe and Mg line ratios on
radiation between 912 -- 1600 \AA.  The H$\alpha$ detection experiments
push the technological limits of ground-based imaging, but the
absorption line approach is technically feasible for HST (with GHRS,
STIS, or COS), given appropriately bright targets such as 3C~232.

In \S~2 of this work we describe our observations. In \S~3 we describe
the reanalysis of the Stocke et al. (1991) \ha\ images and the limit
they impose on \fesc.  In \S~4 we interpret the GHRS spectra and
constrain \phion\ and the shape of the extragalactic spectrum. In \S~5
we review our results for \phion\ and \fesc\ and place our models in the
context of these and other observations.

\section{OBSERVATIONS}
\subsection{Narrowband H$\alpha$ Images}

To constrain the ionizing flux from the local contribution of NGC 3067,
we reanalyzed the H$\alpha$ images presented in Stocke et al. (1991).
These images were obtained at the KPNO 2.1 m telescope in July 1989.
The dataset consisted of two 20-min exposures through a narrow-band
interference filter centered on the redshifted H$\alpha$ wavelength
($\lambda _{c} = 6606 $ \AA, $\Delta \lambda$ = 75 \AA) and a
corresponding ``off-band'' image through a similar interference filter
centered 400 \AA\ longward of the first and chosen to avoid the
redshifted \ion{S}{2} $\lambda\lambda$6719,6731 doublet.  The images
were flat-fielded and calibrated so that 1 DN in the images equals $1.28
\times 10^{-15}$ erg cm$^{-2}$ s$^{-1}$ incident on the telescope.

Using sky background well away from the location of the absorbing
cloud, we scaled the overall background to identical mean values in the
on-band and off-band images and then subtracted the latter from the
former to produce a ``pure'' H$\alpha$ image of the galaxy.  In the
pure H$\alpha$ image we added up all the detected H$\alpha$ photons in
the immediate region of the galaxy to obtain an integrated H$\alpha$
flux for the galaxy of 0.439 photons cm$^{-2}$ s$^{-1}$ incident at the
observer. If the galaxy lies at the velocity of its \ion{H}{1} 21-cm 
emission, then its distance $d$ = 18.9 h$_{75}^{-1}$ Mpc, and the
total \ha\ photon luminosity is $Q_{H\alpha} = 2 \times
10^{52}~h_{75}^{-2}$ photons s$^{-1}$ (where $h_{75}$ is the Hubble
constant $H_{0}$ in units of 75 \kms\ Mpc$^{-1}$).  This result is used
below (\S~3) to impose a limit on the Lyman continuum escape fraction
\fesc.

\subsection{GHRS Spectroscopy} 
\begin{figure}
\plotone{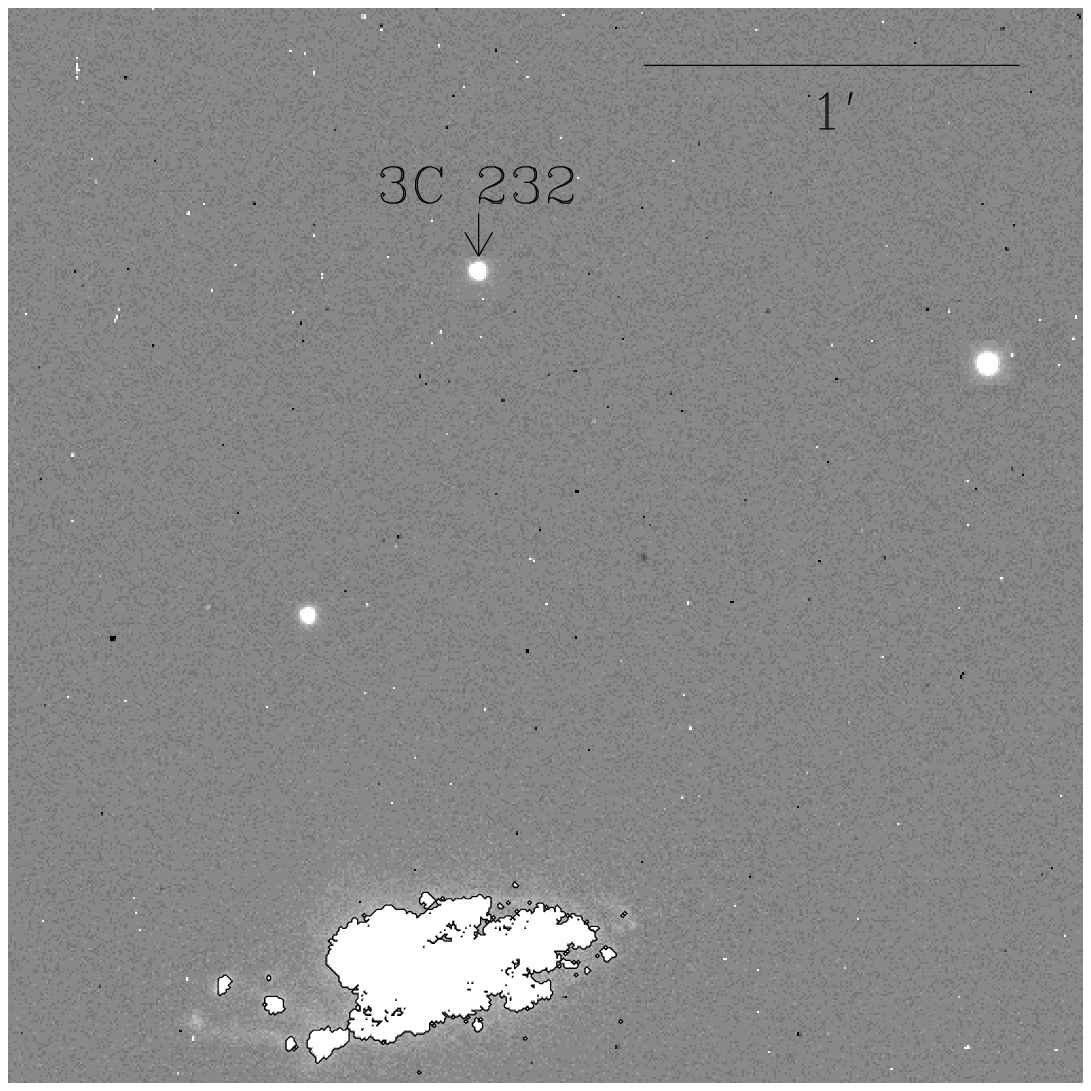} 
\caption{H$\alpha$ image of 3C~232/NGC~3067 field showing the
contour inside which the total H$\alpha$ emission was measured.
The contour lies at 100 DN, or $1.28 \times 10^{-13}$ erg cm$^{-2}$ s$^{-1}$ 
incident on the telescope. 
\label{fig1}}  
\end{figure}

The extended gas cloud associated with NGC~3067 is intercepted by the
sightline to the quasar $\sim 8~h_{75}^{-1}$ kpc above the disk of the
galaxy (Figure~\ref{fig1}; Stocke et al.~1991; Carilli \& van
Gorkom~1992).  We observed 3C 232 to measure the absorption lines in
the quasar spectrum that arise in the cloud.  The galaxy lies at
redshift $z$ = 0.00474 (Carilli \& van Gorkom 1992) and the quasar lies
at $z_{em}$ = 0.532 (Hewitt \& Burbidge 1993). These observations were
carried out during HST Cycle 6 in 1996 using the GHRS with the G270M
grating (R~$\sim$~28,000),the GHRS Large Science Aperture (LSA), and
post-COSTAR optics.  Five grating carousel positions were used to
center the spectra on the lines of interest. A journal of these
observations appears in Table~\ref{table1}. In addition to measuring
absorption in the halo gas of NGC~3067, we measured lines of the same
ionization species in the interstellar medium of the Milky Way and
detected a \ion{C}{4} absorption complex that is apparently intrinsic
to the quasar. Results of these latter two parts of the investigation
will be published separately.

\begin{table} 
\dummytable\label{table1} 
\end{table} 

The data were calibrated with the standard STSDAS/GHRS pipeline and the
most recent available reference files (as of April 1998). With this
software, we subtracted the background, calibrated the wavelengths and
fluxes, and co-added the spectra.  We subtracted a polynomial
background continuum model to avoid the oversubtraction associated with
readout of data over the South Atlantic Anomaly (De La Pena \&
Soderblom 1997).  We used a custom error-propagation routine to
calculate the errors pixel-by-pixel with standard statistical methods.

An uncertainty in the zero-point wavelength of the individual exposures
is introduced into the GHRS data by the uncertainty in the position of
the target within the LSA. The aperture subtends 1.74$''$ and 2.5
diodes on the detector, which corresponds to 0.96 \AA\ with G270M and
causes a 17 \kms\ arcsec$^{-1}$ shift in the observed velocities. The
actual shift could be as low as 5 \kms, given the typical 0.1$''$
accuracy in pointing of the telescope. Since we observe
close velocity coincidence between \ion{Mg}{2} absorption and the
\ion{Na}{1} components reported by Stocke et al.~(1991), and because 
our results are not dependent on the precise wavelength calibration, 
we can tolerate velocity uncertainty at this level. 

\begin{figure}
\plotone{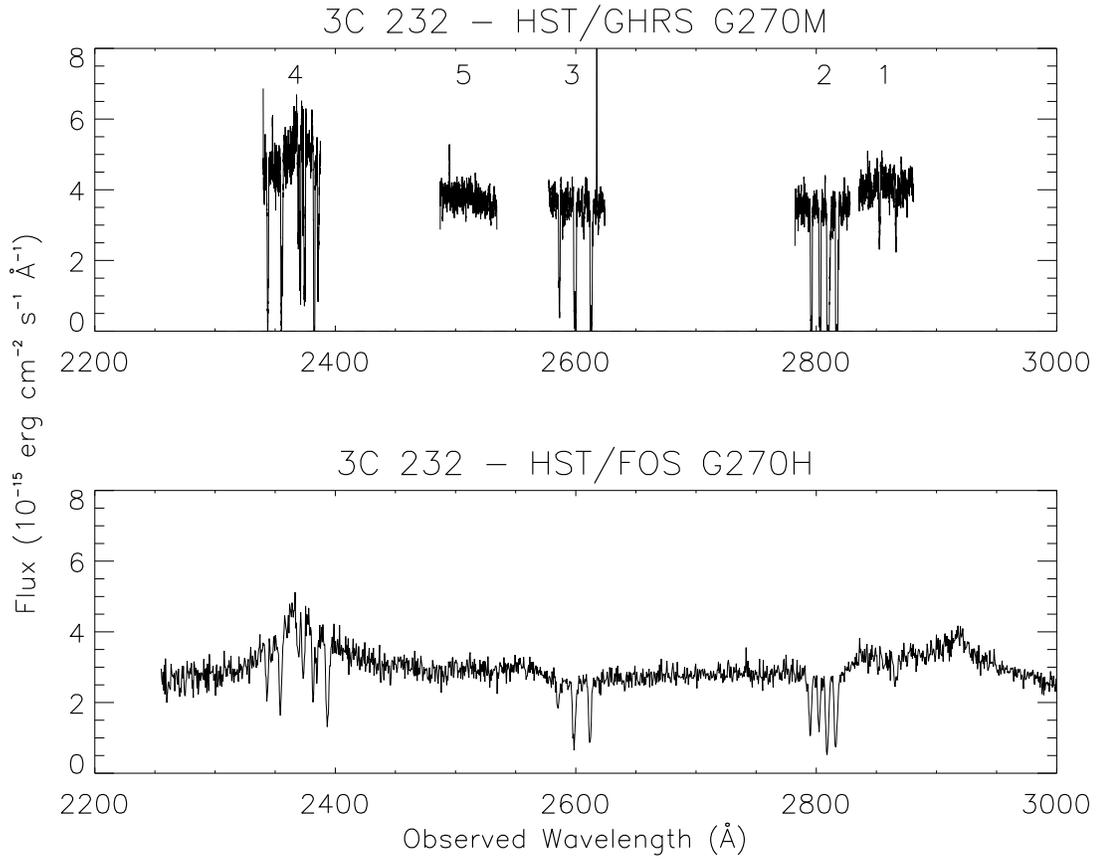} 
\caption{GHRS/G270M and FOS/G270H spectra of 3C~232. The five
GHRS grating settings in the upper panel correspond to the entries
in Table~\ref{table1}.\label{fig2}}
\end{figure}

\begin{figure}
\plotone{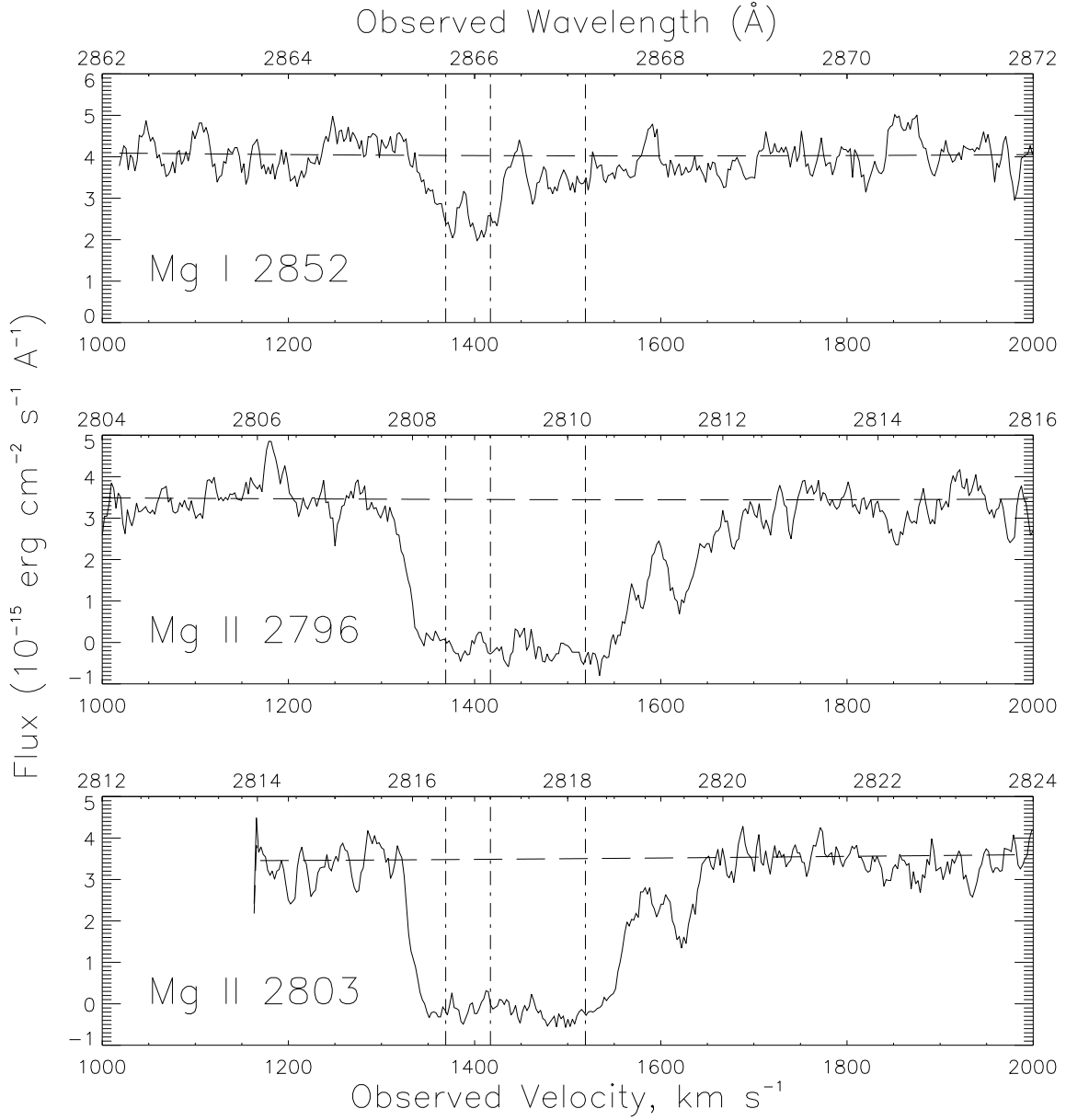} 
\caption{Mg lines observed in 3C 232 plotted versus observed
heliocentric velocity.  Dotted lines show the fitted continuum.
Vertical dashed lines show velocities of three components observed by
Stocke et al. (1991) at V$_{hel} =$ 1369, 1417, and 1530 km
s$^{-1}$.\label{fig3}}
\end{figure}

\begin{figure}
\plotone{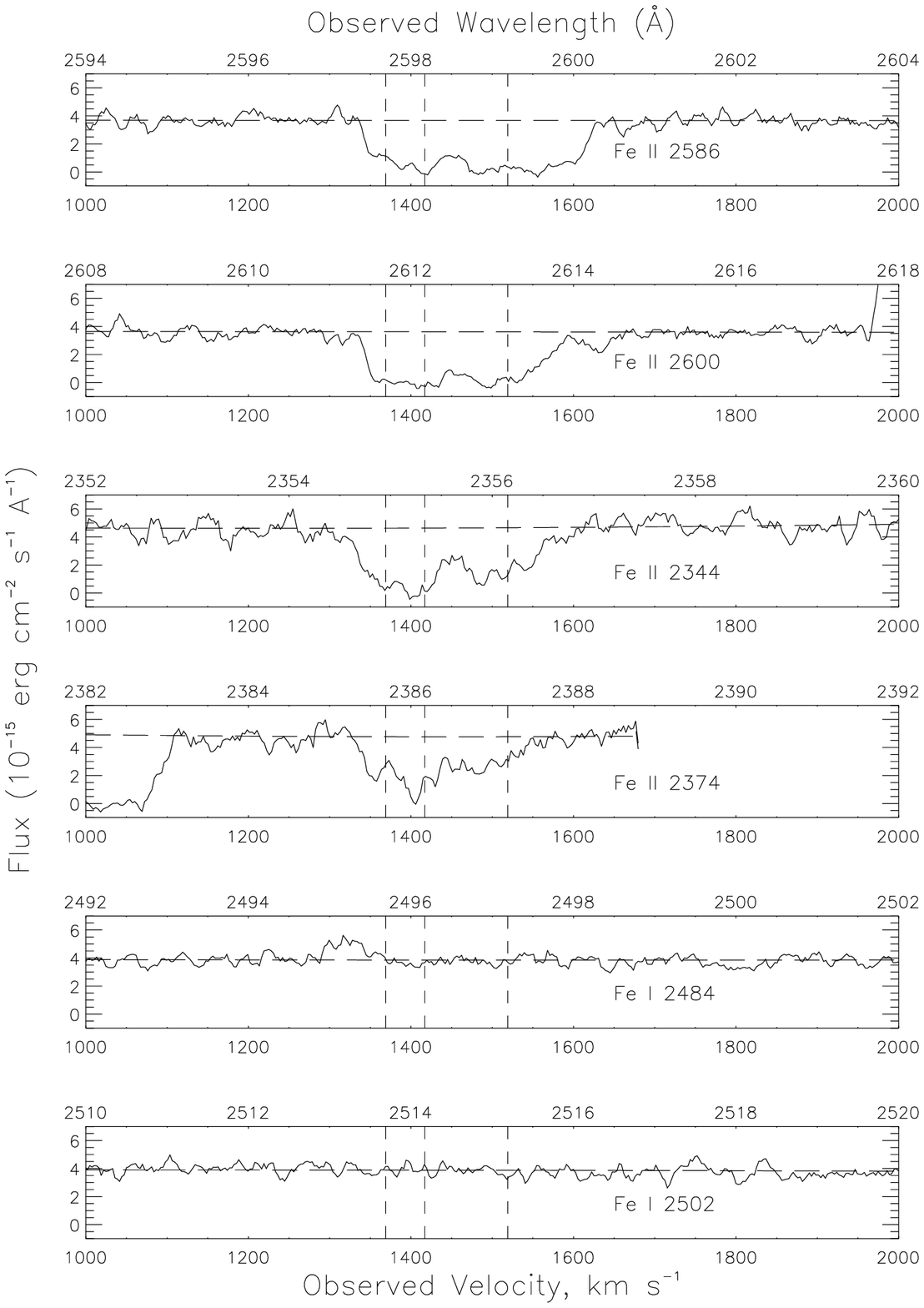} 
\caption{Same as Figure~\ref{fig2} for \ion{Fe}{1} and \ion{Fe}{2}.\label{fig4}}
\end{figure}

The complete, reduced spectra appear in Figure~\ref{fig2}, together
with the previous FOS spectra of 3C 232 from the HST Key Project
(Jannuzi et al.~1998) Close-ups of the lines detected in the new
GHRS/G270M spectra appear in Figures~\ref{fig3} and~\ref{fig4}, plotted
in observed velocity space to show their coincidence\footnotemark[1].
The oversampled data have been smoothed to instrumental resolution (0.1
\AA) by a 5-pixel sliding boxcar filter.  We judge the statistical
significance of absorption features by imposing a $3\sigma$ cutoff
depth for positive identification at their predicted velocities.  A
summary of the detected lines appears in Table~\ref{table2}.

\begin{table}
\dummytable\label{table2}
\end{table}

The most striking feature of the new spectra is the 30\% increase in
the mean level of the quasar continuum between the two epochs (FOS in
1992 and GHRS in 1996). Because this difference exists even after both
spectra are calibrated with the best available calibration files, we
attribute this increase to variability in the quasar.  Bruhweiler et
al. (1986) reported variability in 3C~232 on time scales as short as a
day. The level of the continuum affects our measurements minimally, and
the issue of variability is left for a future discussion of intrinsic
quasar features.

Stocke et al. (1991) reported the detection of three distinct velocity
components in \ion{Na}{1} and \ion{Ca}{2} absorption in this line of
sight.  These components lie at $V_{hel}$ = 1417~\kms~(system 1),
1369~\kms~(system 2), and 1530~\kms~(system 3).  System 1 corresponds
to the velocity, $V_{HI}$ = 1420~\kms, of the cold absorbing \ion{H}{1}
gas reported by Carilli \& van Gorkom (1992).  The majority of the
observed lines in our GHRS spectra are saturated, which erases almost
all detailed information about velocity components.  However, we note
velocity coincidence with the Stocke et al. (1991) components in almost
all lines. In \ion{Fe}{2}, \ion{Mg}{1}, and \ion{Mg}{2} we detect
absorption at the velocities of at least one of the these three
components, and in most cases all three (Figures~\ref{fig3}
and~\ref{fig4}).  In \ion{Mg}{1} $\lambda$2853 we detect absorption at
the velocities of system 1 and system 2.  Strong absorption
corresponding to all three components is detected in the saturated
\ion{Mg}{2} $\lambda \lambda$2796,2804 doublet.  All three components
are detected in \ion{Fe}{2} $\lambda$2586, $\lambda$2600,
$\lambda$2344, and $\lambda$2374, but no lines of \ion{Fe}{1} are
seen.  The non-detection of \ion{Fe}{1} has important implications for
our estimates of $\Phi _{ion}$.  

\footnotetext[1]{A small (+2 \kms) heliocentric correction has been
applied to the observed GHRS data based on the position of 
the target.}  

With the goal of placing conservative error bars on our measurements,
we fitted the continuum on a line-by-line basis with the maximum
possible number of continuum points.  The continuum points were fitted
with Legendre polynomials, which allow the addition of higher-order
terms to the fit without changes in the lower-order coefficients. In
all cases, low-order polynomials were used ($n \leq$ 4).  No higher
terms were added after the $\chi^{2}$ per degree of freedom changed by
less than 5\%.  We measured equivalent widths with two methods. For the
saturated lines, we integrated the line under the fitted continuum
without attempting to fit the profile. We fitted the unsaturated lines
with single or double Gaussian components and calculated equivalent
widths from the best-fit parameters. The tabulated values are the total
equivalent width for each line taken from the best available
measurement.  Errors were calculated according to the comprehensive
scheme outlined in Sembach \& Savage (1992, their Appendix A). This
method estimates errors both from the fitting of the continuum and from
the statistical errors propagated through the calibration, and it
accounts for all sources of statistical uncertainty.

The spectrum at setting 5 would contain \ion{Fe}{1} $\lambda$2484 and
$\lambda$2502 in absorption at the positions indicated in
Figure~\ref{fig4}.  There are no easily distinguishable absorption
features in this spectrum, and there are no features at all deeper than
$2\sigma$.  We have calculated upper limits on the column density of
\ion{Fe}{1} based on $3\sigma$ upper limits on the equivalent widths of
these undetected lines. These $3\sigma$ limits are used in all
subsequent analysis.

Because we detect absorption at velocities corresponding to the three
components detected by Stocke et al. (1991), we argue that the simplest
interpretation of these data is that all three components exist, but
are too high in column density to appear distinctly separate in
\ion{Fe}{2} and \ion{Mg}{2}.  Carilli \& van Gorkom (1992) reported
that the width of the neutral \ion{H}{1} 21-cm absorption line is 4.7
\kms.  We note that 21-cm absorption measures N(\ion{H}{1}) / $T$, and
thus is most sensitive to cold gas.  If this line width is purely
thermal, then Fe and Mg should exhibit line widths correspondingly
lower in proportion to their atomic masses.  Because we see lines in Fe
and Mg that are $\sim 40$ \kms\ broad, we must attribute this
broadening in resolved lines to warm gas, velocity components, or high
column density, which accentuates the high velocity tails of the
particle distribution.

Because we interpret the \ion{Mg}{2} and \ion{Fe}{2} absorption as
occurring in three distinct components, we should assume three
components when setting limits on \ion{Fe}{1} absorption.  We impose a
3$\sigma$ limiting equivalent width of 24 m\AA\ on an individual
component of \ion{Fe}{1}.  It is possible that the smallest column
detectable in a single component occurs at each of the three
velocities, and that all three are undetected.  Therefore the 3$\sigma$
upper limit on \ion{Fe}{1} column density is 9 times the 1$\sigma$
limiting equivalent width.  The strongest upper limit on
$N$(\ion{Fe}{1}) is set by this limiting equivalent width applied to
the strong \ion{Fe}{1} $\lambda$2484 line ($f$~=~0.5569). We note that
this method of setting the ionization balance holds independent of the
explanation for the broad velocity structure of absorption discussed
above because these explanations apply equally well to \ion{Fe}{1} and
\ion{Fe}{2}.

\begin{figure}
\plotone{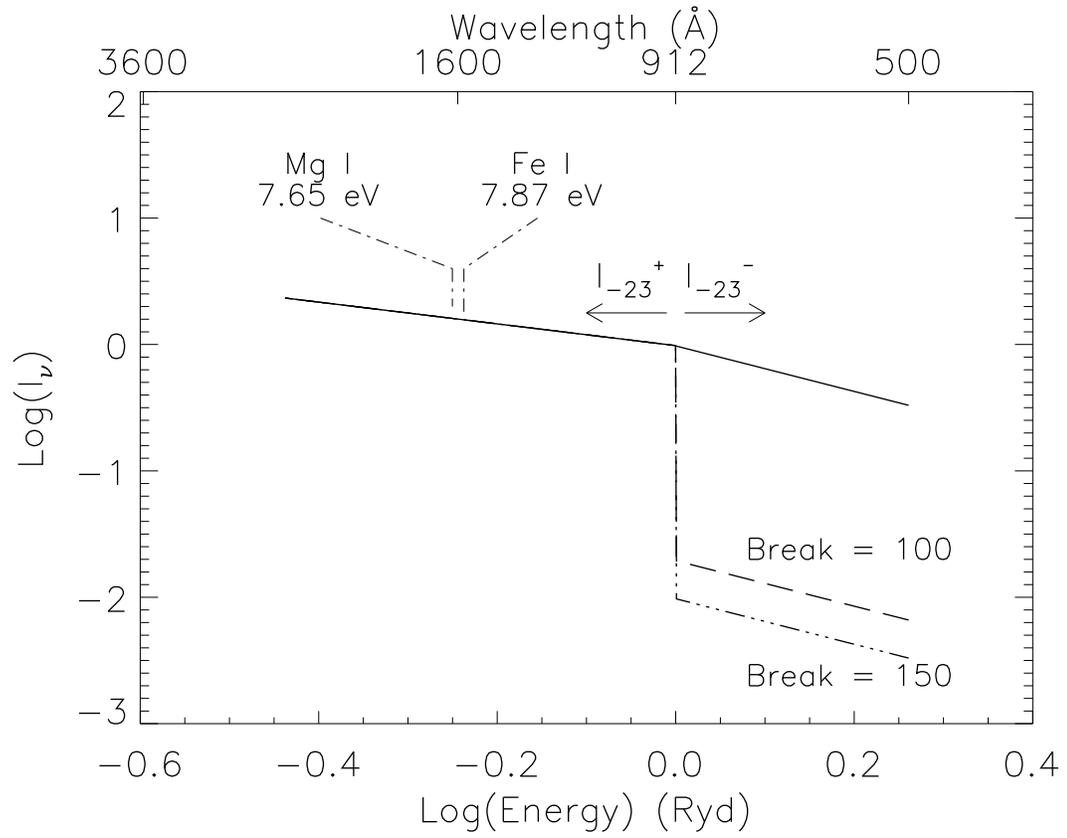} 
\caption{Model spectra: The solid line is from Zheng et al.~(1997),
modified to place the break in spectral index from $\alpha_{s}$ to
$\alpha_{FUV}$ at 1 Ryd. The dashed and dashed-dotted lines represent the
same spectrum with breaks at 1 Ryd as noted.\label{fig5}}
\end{figure}

Using the measurements and limits in Table~\ref{table2},  we construct two critical
ionization ratios \ffe\ ~=~N(\ion{Fe}{1})~/~N(\ion{Fe}{2}) and
\fmg\ ~=~N(\ion{Mg}{1})~/~N(\ion{Mg}{2}). From \ion{Mg}{1}
$\lambda$2853 and \ion{Mg}{2} $\lambda$2804 we derive \fmg\ $\leq
0.0175$, and from \ion{Fe}{1} $\lambda$2484 and \ion{Fe}{2}
$\lambda$2374 we infer \ffe\ $\leq 0.0044$.  These ratios are used
below (\S~4.1) to constrain the ionizing spectrum incident on the
cloud.  Our limit on \phion\ depends on the ratio of total absorption
in \ion{Fe}{1} to the total absorption in \ion{Fe}{2}, and to a lesser
extent on the ratio of \ion{Mg}{1} to \ion{Mg}{2}. Because the
ionization edges of Mg I and Fe I lie at 7.646 eV and 7.87 eV,
respectively, these ratios are sensitive to the shape of the spectrum
in the FUV band (Figure~\ref{fig5}).

\section{IONIZING RADIATION FROM NGC 3067}

The narrow-band \ha\ images of NGC~3067 measure the total \ha\ photon
emission rate from the galaxy NGC 3067, $Q_{H\alpha}$ = $2 \times
10^{52} h_{75}^{-2}$ s$^{-1}$.  If we assume, per case-B recombination
theory, that these regions emit one H$\alpha$ recombination photon per
2.4 ionizing photons at $T = 20,000$ K (Osterbrock 1989), then the
ionizing photon production rate of NGC 3067 is $Q _0 = 5 \times
10^{52}~h_{75}^{-2}$ s$^{-1}$.  This result implies \phion\ $= 4 \times
10^{6}$ \hsc$^{-2}$ \units\ if the \ion{H}{2} regions in NGC 3067 are
assumed to be a point source and no opacity is included in the galactic
disk or halo.  We account for geometric effects by directly integrating
the individual pixels in the image, attenuated by their projected
distances from the quasar sightline.  This method increases our
estimate of the galactic ionizing flux by 5\% but still does not
account for the full ionizing flux in the unlikely event that the the
edge-on disk of the galaxy is optically thick to H$\alpha$.

Stocke et al. (1991) used these same images to calculate an upper limit
to the \ha\ emission from the cloud. From a simple, slab-geometry
analytical photoionization model of the cloud, they derived
\phion\ $\geq 8.35 \times 10^{4}$ \units.  For a smooth spectrum, our
measurement \phion\ $= 4 \times 10^{6} h_{75}^{-2}$ \units\ and the
Stocke et al.~(1991) limit sets \fesc\ $\leq$ 0.02.  Because the two
numbers in the ratio derive from the same images, the errors in the
ratio are dominated by statistical fluctuations in the image and not by
calibration errors or systematic uncertainties associated with
combining measurements by different instruments.

The inferred ionizing flux at the cloud due to the ionizing sources in
the galaxy could dominate any extragalactic contribution at 8 kpc from
the disk, were the galactic disk and halo transparent to ionizing
radiation. It is known that galaxies exhibit Lyman breaks due to the
opacity of \ion{H}{1} in stellar atmospheres and in the galactic ISM.
However, the fraction of Lyman continuum photons that escape galaxies
and contribute to the extragalactic background, as parametrized by
\fesc, is poorly constrained (see \S~1).  The limit \fesc\ $\leq 0.02$
is more stringent than those of Leitherer et al. (1995) and lies 
below the limits Hurwitz et al.~(1997) inferred from reanalysis of the
same data.  Our robust upper limit on \fesc\ leaves open the
possibility that the portion of the galaxy's ionizing radiation that
escapes the halo is negligible from the point of view of the absorbing
cloud. The assumption that the cloud is illuminated by a purely
extragalactic spectrum permits a lower limit to the intensity of the
ionizing radiation field, as discussed below in \S~4.

\section{PHOTOIONIZATION MODELS OF THE ABSORBING CLOUD}

Detailed photoionization models permit comprehensive exploration of the
effects of various ionizing continua and total ionizing fluxes on a
model cloud that can be compared against the observations.  An
optically thin photoionization model fails to describe this cloud
accurately and to predict the ionization balance of its metals, in
large part because there is no easy way to predict {\em a priori} the
electron fraction in the cloud.  The cloud has a hydrogen column
density sufficient to shield its interior from radiation at $h\nu \geq
13.6$ eV, so that a treatment of radiative transfer within the cloud is
necessary to predict metal-line ratios.  As a result, we simulated the
extended \ion{H}{1} halo of NGC~3067 with the Cloudy code from G.
Ferland (version 90.04, Ferland 1996).

In all cases, the model cloud was a plane-parallel slab of thickness
$D_{HI} \sim$ 6.5$h_{75}^{-1}$ kpc, consistent with the angular size of
the cloud set by the limiting column density contour at $N_{HI} =
2\times 10^{19}$ cm$^{-2}$ from Carilli \& van Gorkom (1992).  The
model was integrated in depth from its illuminated face until the
column density of neutral hydrogen reached the peak value $N_{HI} = 8
\times 10^{19}$ cm$^{-2}$, measured by Carilli \& van Gorkom (1992).
In all models, the total hydrogen density $n_{H}$ was assumed constant
and was varied to produce the observed $N_{HI}$ within a physical size
comparable to $D_{HI}$.  The assumption that the cloud is illuminated
from one side is consistent with both of the models proposed:  either
the cloud is illuminated by a purely extragalactic radiation field and
is shielded from the galaxy by the inner halo gas and dust, or the
nearby galaxy dominates the spectrum between 912 -- 1600 \AA\ and the
extragalactic contribution is negligible. Only the shape and intensity
of the continuum varies between models.

Because the ionization edges of Mg I and Fe I lie at 7.65 eV and 7.87
eV, respectively, radiation at energies between 0.56 and 1.00 Ryd
strongly influences the calculated ionization ratios.  We employ two
generic spectra to explore the dependence of the ionization ratios
\ffe\ and \fmg\ on the extragalactic UV radiation (Figure~\ref{fig5}).
One of these assumes a continuous power-law spectrum representative of
extragalactic radiation dominated by AGN (\S~4.1). The other assumes
that local and possibly integrated galactic contributions are important
and contains a jump at the \ion{H}{1} ionization threshold (\S~4.2).

We employ five constraints on the model cloud and its incident
radiation field.  The ionization ratios \ffe\ and \fmg\ are highly
sensitive to FUV radiation shortward of 1600 \AA\ at the cloud. We use
the critical value \ffe\ $= 0.0044$ to constrain the FUV background as
well as the extragalactic ionizing background \phion.  The
\fmg\ constraint is satisfied by all models that meet the
\ffe\ constraint. The measured column of the \ion{Mg}{1} $\lambda$2853
lines constrains the total FUV flux between 912 -- 1600 \AA, and with
an assumption of \phion, this line can be used to constrain the 13.6 eV
break in the spectrum. The observed $N_{HI}$ and $D_{HI}$ constrain the
size and density of the model clouds.

\subsection{Models with Continuous Power-law Spectra}

The case of a purely extragalactic background is the simplest limiting
class of model that can account for the ionization ratios in the
absorbing halo cloud.  This model assumes that the cloud is shielded
from the ionizing radiation of the galaxy and that the extragalactic
background is composed of AGN spectra in the power-law form given by
Zheng et al.~(1997). This composite spectrum has a spectral index $\alpha
_{s} = 1.8 \pm 0.3$ shortward of 1050 \AA, while the index changes
to $\alpha _{FUV} = 0.86 \pm 0.01$ for $\lambda \geq$ 1050 \AA. While
this spectrum can be taken as the best available representation of the
QSO-dominated extragalactic spectrum, we modify it in two ways to
include the possibility that the power-law is dominated by normal
galaxy contributions. First, we move the knee where $\alpha$ changes
from 1050 \AA\ to 912 \AA. In practice this change is unnoticeable in the
ionization ratios. Second, we include a break at the Lyman limit in
some models, as discussed in \S~4.2.

\begin{figure}
\plotone{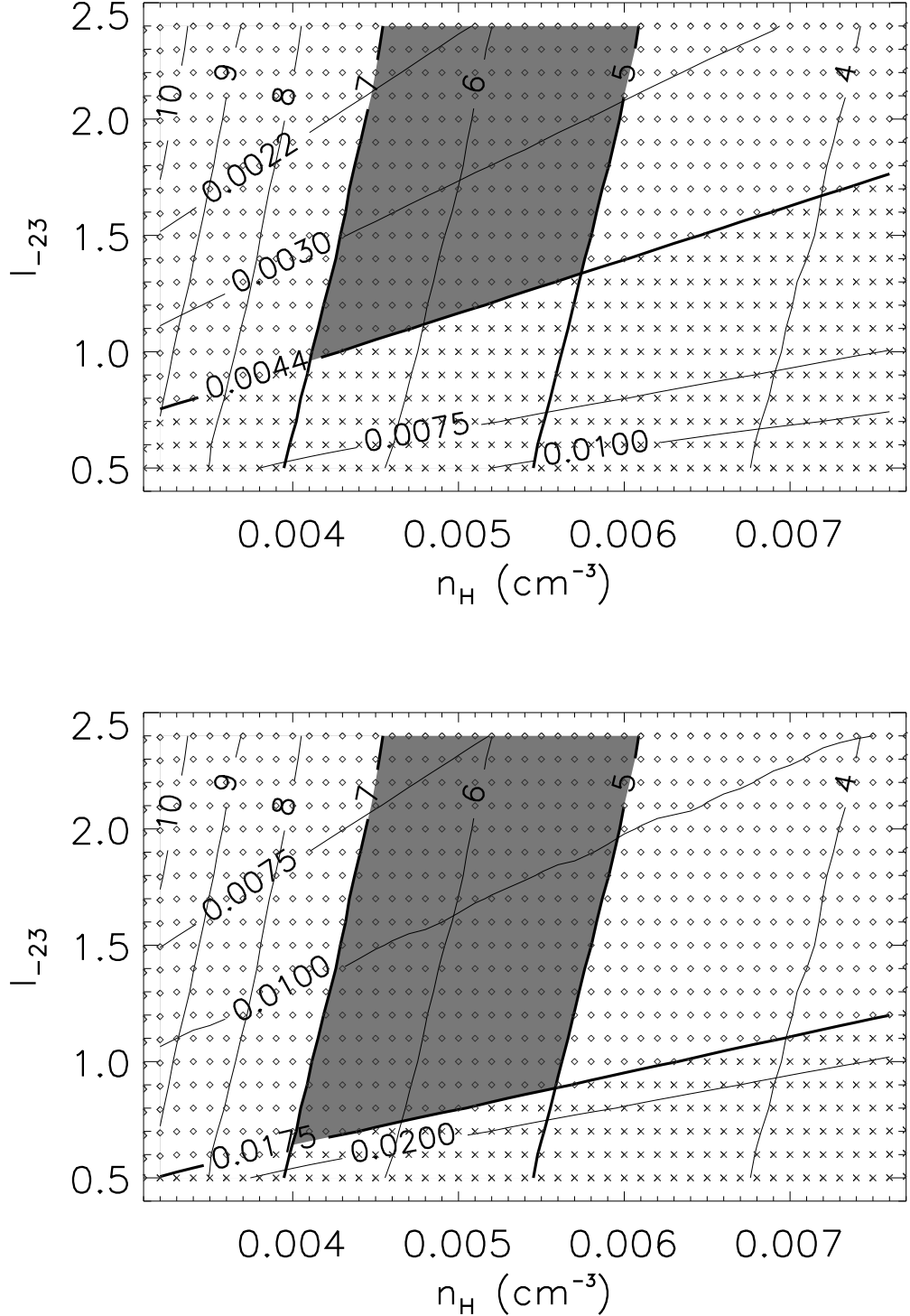} 
\caption{Upper panel: Smoothed parameter space for Fe lines showing
contours of constant ratio \ffe\ $=$ N(\ion{Fe}{1})~/~N(\ion{Fe}{2})
between 0.0034 and 0.0100 and constant cloud diameter $D_{HI}$ in kpc.
A featureless power-law extragalactic spectrum with $\alpha _{s} = 1.8$
is assumed.  $I_{-23}$ is the intensity of the model spectrum at the
Lyman limit in units of $10^{-23}$ erg cm$^{-2}$ s$^{-1}$ Hz$^{-1}$
sr$^{-1}$.  Lower panel: Same for \fmg\ $=$
N(\ion{Mg}{1})~/~N(\ion{Mg}{2})  ratio with \fmg\ ranging between 0.010
and 0.025.\label{fig6}}
\end{figure}

For a suite of model clouds illuminated by the modified composite QSO
spectrum, the input values of $I_{-23}$ and $n_{H}$ were varied to
produce a contour plot of the parameter space (Figure~\ref{fig6}).  The
contours represent constant values of the ratios \ffe\ or \fmg.  The
tightest constraint on $I_{-23}$ is the contour \ffe\ $= 0.0044$, set
by the ratio of the tightest upper limit on N(\ion{Fe}{1}) to
N(\ion{Fe}{2}) from \ion{Fe}{2} $\lambda$2374.  We take the acceptable
range of parameter space to be the shaded area, based on our limits on
\ffe\ and the apparent diameter $D_{HI} \sim$ 6.5 $h_{75}^{-1}$ kpc of
the cloud from the \ion{H}{1} emission map of Carilli \& van Gorkom
(1992).  Because the \ion{Mg}{2} $\lambda\lambda$2796,2804 lines are
highly saturated, the constraint \fmg\ $\leq 0.0175 $ on \phion\ is
satisfied by all models that meet the \ffe\ constraint.

This constant-density model requires that \phion\ $\geq$ 2600
$\,$cm$^{-2}\,$s$^{-1}$ and $I_{-23} \geq 1.0$ for $\alpha _{s} =
1.8$.  We judge models in the parameter space to be acceptable if they
lie above the contour corresponding to the critical value of \ffe.
Sources of uncertainty in the ratio, and therefore in the contours, are
discussed below (\S~4.3).

Our limit is lower than all the upper limits on \phion\ imposed by
measurements and non-detections of H$\alpha$ (Vogel et al. 1995;
Donahue et al. 1995; Stocke et al. 1991; Kutyrev \& Reynolds 1989), and
it falls near the lower end of estimates of \phion\ from \ion{H}{1}
galactic disk edges (Maloney 1993; Dove \& Shull 1994). Our limit lies
within the large range of \phion\ derived by Kulkarni \& Fall (1993)
from the Ly$\alpha$ forest proximity effect at $\langle z \rangle$ =
0.5.  However, direct comparison of our limit to these previous
measurements is valid only in the case that the local contribution of
\g\ is negligible. If \phion\ is dominated by the ionizing radiation
from the galaxy, the extragalactic ionizing background cannot be
constrained.


Although the cloud appears to have a roughly spherical profile when seen
in projection, we have no reason to expect that it has any particular
density profile.  Indeed, a spherical cloud of constant density 
exhibits enhanced column density near its projected center.
Furthermore, in this case, uncertainties in the density profile are
dominated by the larger uncertainty about the source of the ionizing
radiation.  Because centrally condensed models require higher values of
\phion\ to match the observed \ffe\ and \fmg, our lower limit is not
sensitive to the assumed profile.  The limit is also insensitive to
density clumps that may lie hidden in VLA maps with $\sim$ 50$''$
resolution. Because we cannot constrain the size and filling factor of
these clumps, and given the relative insensitivity of our limit on
\phion\ to this subtlety, more complicated models are unjustified.

\subsection{Models with Lyman Breaks} 

The second class of models incorporate both QSO and starburst
contributions to the ionization of the absorbing cloud.  These models
keep the power-law form of the spectrum but contain a ``break'' at 13.6
eV, such that the continuum longward of the Lyman limit is enhanced
relative to the H-ionizing continuum.  Recent theoretical work by Shull
et al.~(1999) demonstrates that if \fesc\ $\sim$ 0.05, the contribution
from galaxies to the extragalactic ionizing background may match the
empirically evaluated contribution of AGN.  Because an intrinsic Lyman
break is introduced by \ion{H}{1} in stellar atmospheres and by the
galactic ISM, galaxies may contribute far more to the extragalactic FUV
(912 -- 1600 \AA) background than they add to the ionizing continuum.
As a result, an incident spectrum with a break at 13.6 eV may represent
the extragalactic radiation background better. This is particularly
relevant for \ffe\ and \fmg. The same general argument for a Lyman
break may be applied specifically to the contribution to the incident
radiation by the local starburst galaxy \g. We return to this point
below.

Figure~\ref{fig5} illustrates the schematic spectra we assume are
incident on the model cloud. The solid curve represents the minimum
level of the ionizing background from \S~4.1 required to satisfy the
limits on \ffe\ and \fmg. If the flux at energies below 1 Ryd is
enhanced by at least a factor of two over that of the ionizing
spectrum, the observational constraints \ffe\ and N(\ion{Mg}{1}) may be
satisfied independent of the level of ionizing radiation by appropriate
choice of FUV flux.

\begin{figure}
\plotone{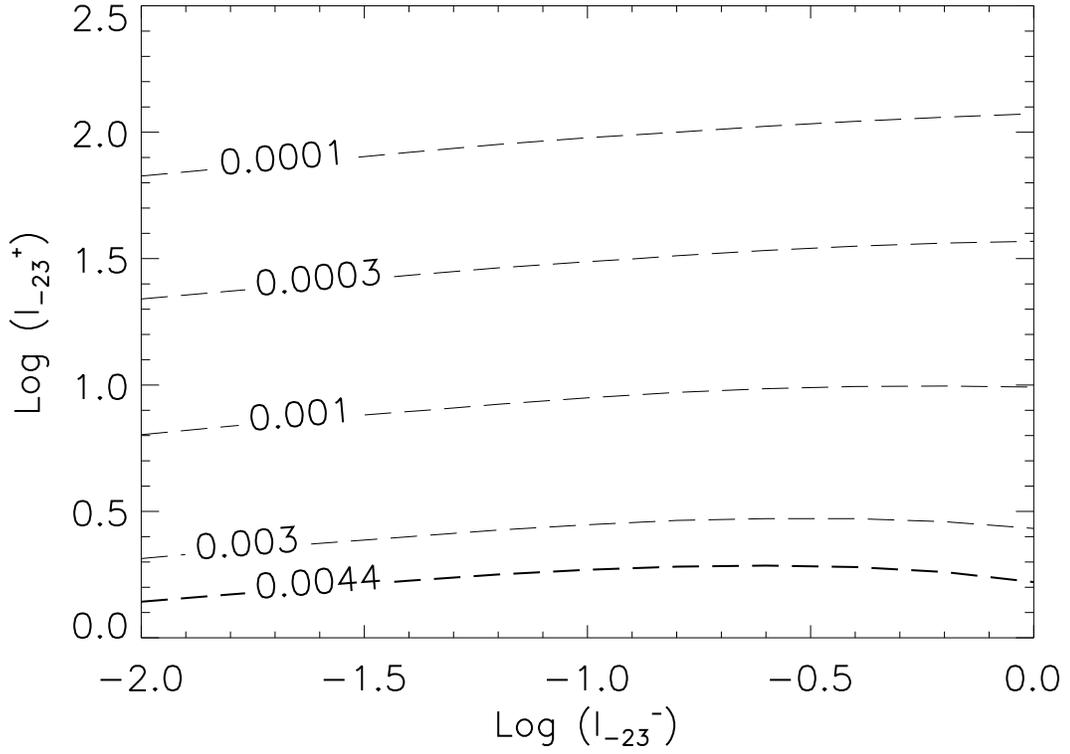} 
\caption{Contour plot of \ffe\ with varying $I^{+}_{-23}$ and
$I^{-}_{-23}$ (longward and shortward of the Lyman limit,
respectively).  For a given value
of $I_{-23}^{+}$, the ionization ratio does not constrain the value of
$I_{-23}^{-}$, because our upper limit to \ffe\ allows one to choose
any point above the \ffe\ = 0.0044 contour.\label{fig7}}
\end{figure}

\begin{figure}
\plotone{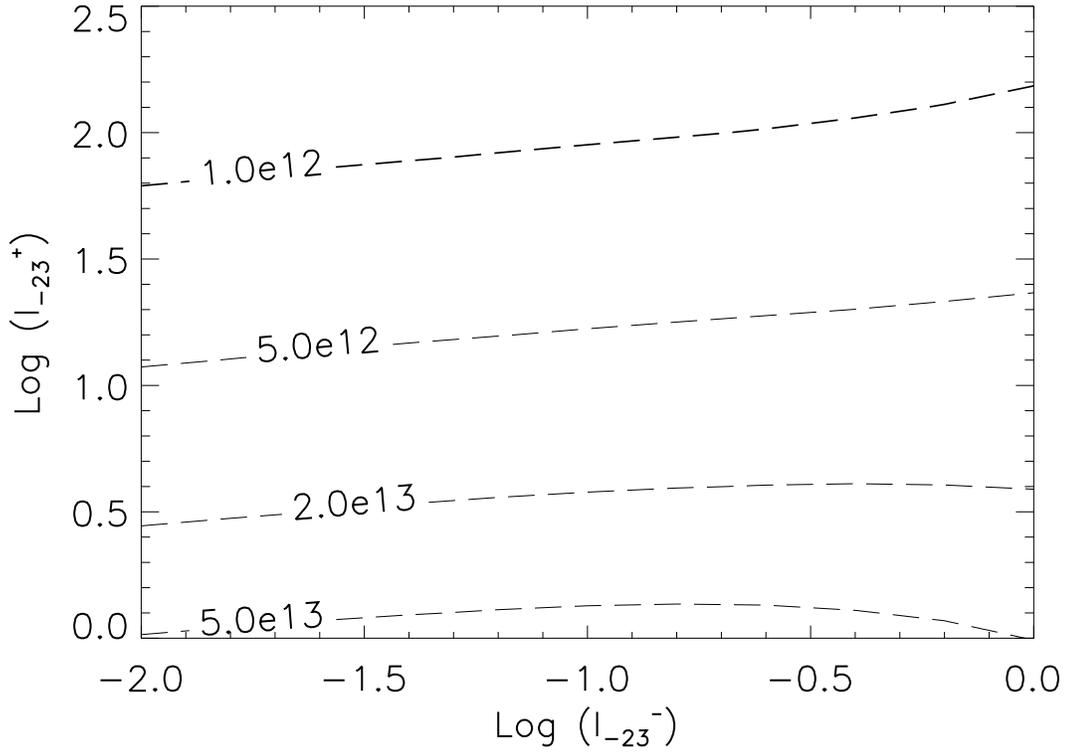} 
\caption{Contour plot of N(Mg I) with varying $I^{+}_{-23}$ and
$I^{-}_{-23}$ (intensity longward and shortward of the Lyman limit,
respectively).  For an assumed $I_{-23}^{-} = 1.0$, or \phion\ $= 2600$ \units,     
$I^{+}_{-23}$ translates directly into the break expected in the
incident spectrum. In this case, the presence of the
\ion{Mg}{1} $\lambda 2853$ line constrains the break to 130 -- 150 for
$I_{-23}^{-} = 1.0$. For vanishing \phion, the break becomes
arbitrarily large.\label{fig8}}
\end{figure}

Figures~\ref{fig7} and~\ref{fig8} illustrate the use of the quantities
$I^{+}_{-23}$ and $I^{-}_{-23}$ to represent the levels of incident
radiation just longward and just shortward of the Lyman limit,
respectively.  Although there is likely not a pure step function break
in the spectrum at the Lyman limit, these quantities schematically
represent overall levels of FUV (912 -- 1600 \AA) radiation
($I^{+}_{-23}$) and H-ionizing radiation ($I^{-}_{-23}$). As
Figure~\ref{fig7} shows, if $I^{+}_{-23} \geq 2$, constraints on
\ffe\ are satisfied independent of $I^{-}_{-23}$, and, by implication,
independent of \phion. A similar conclusion holds for N(\ion{Mg}{1}).
Figure \ref{fig8} shows the levels of $I^{+}_{-23}$ and $I_{-23}^{-}$
allowed by the N(\ion{Mg}{1}) constraint.  If $I_{-23}^{-}$ is assumed,
the allowed values can be expressed in terms of an allowed ``break'' in
the ionizing spectrum.

If we fix \phion\ $=$ 2600 \units, we can vary the flux longward of 912
\AA\ until the \ion{Mg}{1} line in the model cloud is no longer
observable. For this calculation, we assume a 3$\sigma$ lower limit on
N(Mg~I) $\leq 1 \times 10^{12}$ cm$^{-2}$ and no depletion.  The
greatest break that accommodates the presence of Mg I absorption is a
factor $\sim$ 150, assuming the same range of densities as in
Figure~\ref{fig6}.  Figure~\ref{fig8} shows contours of constant
N(\ion{Mg}{1}) in the model cloud for a varying break. The bold contour
represents the minimum measured N(\ion{Mg}{1}); thus, for \phion $\sim
2600$ \units, $I_{-23}^{-} \sim 1.0$ and the break is constrained to
$\sim 150$.  Higher breaks are of course permissible for vanishing
\phion, and they do not violate the \ffe\ constraint
(Figure~\ref{fig7}).

Direct measurements of the extragalactic FUV background longward of 912
\AA\ are very difficult, due to the need to subtract out local
contributions to the observed radiation (Bowyer 1991; Henry 1991;
Witt et al.~1997). In general, these highly uncertain estimates are
consistent with an upper limit to the break of order 200 -- 300,
assuming \phion\ $=$ 2600 \units.  Armand et al. (1994) suggest an
integrated contribution from galaxies at 2000 \AA\ consistent with a
break of $\sim$ 50 -- 150, again assuming \phion\ $=$ 2600 \units.  The
presence of \ion{Mg}{1} in the \g\ cloud is compatible with these
estimates of the extragalactic FUV background.

It is more difficult to argue that the local 912 -- 1600
\AA\ contribution of \g\ may be neglected than it is to dismiss its
contribution to the ionizing radiation incident on the cloud.  Our
limit \fesc\ $ \leq 0.02$ on the escape of ionizing photons becomes
weaker for these lower-energy photons. As Sutherland \& Shull (1999)
show, synthetic starburst spectra have breaks in flux at 13.6 eV of
$\sim 4$ for young starbursts, and $\sim 10$ for evolved ones. This
result implies upper limits on the break due to \g\ of $\sim 100-250$
even before considering the enhanced escape probability of photons in
this energy range. In the opacity model of Deharveng et al. (1997), the
local contribution of ionizing photons by \g\ may arguably be
neglected, but the neutral hydrogen required to account for the Lyman
continuum opacity does not impede the escape of FUV photons, so that
\g\ may produce a break of $I^{+}_{-23} \sim 6000$. The geometric 
scenario for \fesc\ is compatible with the inclusion of opacity, such
as dust, which can also limit the escape of FUV photons. If dust is the
primary source of FUV and LyC opacity, because \fesc\ $\sim 0.02$, an
optical depth $\tau _{\lambda} ^{dust} \sim 4$ across the 912 -- 1600
\AA\ band is required.  Using a standard selective extinction curve
(Seaton 1979) and assuming that $A(\lambda)$ continues to increase
shortward of 1040 \AA, we derive the limit $A _{V} \leq 1$ mag, where the
upper limit is due to the uncertainty in the exact form of the
extinction curve shortward of 1000 \AA. For the derived limit
\phion\ $\geq 2600$ \units\ we obtain an upper limit to the V-band dust
extinction $A _{V} \leq 2$ mag. Thus, dust opacity could contibute
significant FUV opacity and modest optical extinction and should be
incorporated into either formal model of \fesc.

If the photons longward of 912 \AA\ escape largely unimpeded from \g,
the presence of \ion{Mg}{1} absorption implies that the gas associated
with it must approach density $n_{H} \sim 0.6$ cm$^{-3}$, assuming that
the Mg is present at solar metallicity. At these densities, depletion
may become significant, further increasing the constraints on the
density.  In addition, assuming that the opacity model of \fesc\ is the
more accurate description, the emergent spectrum should be
significantly hardened shortward of the Lyman limit by the thick layer
of \ion{H}{1} that enshrouds the galactic disk.  Our photoionization
modeling reveals that the indicator ratios \ffe\ and \fmg\ are
insensitive to this hardening.  However, the \ion{C}{2} / \ion{C}{4}
and \ion{Si}{4} / \ion{C}{4} ratios in the model cloud show moderate
changes of 0.4 -- 0.9 dex if the input spectrum is hardened by the
thick \ion{H}{1} disk in the opacity model.

\subsection{Sources of Uncertainty in the Models}

Both the local and extragalactic interpretations presented above are
subject to uncertainties in the total amount of gas present in the 3C
232 cloud, the size of the cloud, and the shape of the ionizing
continuum.  It is possible that the peak column density in this cloud
measured by Carilli \& van Gorkom (1992) and assumed in all the cloud
models discussed here is an underestimate of the actual amount of
\ion{H}{1} in this cloud. The original \ion{H}{1} emission maps were
produced at VLA in C array, with a 20$''$ by 30$''$ beam that was
binned to 50$''$ to improve signal-to-noise.  If these observations
suffered from beam dilution, the actual column at the position of the
quasar may peak at a far higher value. Higher average values of the
total hydrogen density $n_{H}$, provided by a higher column density or
by clumping, would increase the lower limit on \phion.  These changes
effectively rule out the regions of parameter space shown above for all
larger columns of \ion{H}{1} and for centrally condensed density
profiles. For example, a column $N_{HI} = 1.6 \times 10^{20}$ \cd
requires that twice the gas be accommodated within the same physical
size, such that the diameter contours in Figure~\ref{fig6} shift to the
right and the limit becomes \phion\ $\geq 6500$ \units.  

For the case of purely extragalactic ionizing radiation, our limits
are insensitive to changes in the slope of the ionizing continuum. The
limit \phion\ $\geq$ 2600 \units\ changes by roughly 10\% for changes
in $\alpha _{s}$ from 1.5 to 2.1 (Zheng et al.~1997).  Because we rely
exclusively on the ratios of different ions of the same elements, our
limits are not sensitive to the abundance or depletions of these
elements in the cloud.

Of course, the greatest uncertainty in these models is the unknown
contribution of the nearby galaxy, \g, to the ionization of the halo
cloud.  As discussed above, we can reasonably assume that little
H-ionizing flux reaches the cloud. However, the apparent lack of
opacity in the 7.6 -- 13.6 eV range makes the metal-ionizing radiation
field emergent from the cloud potentially important. Only a dust
component in the disk and halo of the galaxy would prevent these
photons from escaping the disk. Thus, the uncertainties of selective
extinction in the far-UV are more important than modest upper limits to
A$_V$.

\section{DISCUSSION AND CONCLUSIONS}

Our absorption line spectra, narrow-band images, and extensive
photoionization modeling of the 3C~232/NGC~3067 system allow us to
limit the ionizing flux at the cloud to \phion\ $\geq 2600$ \units\ if
the ionizing spectrum is continuous.  This limit is consistent with
non-detections of H$\alpha$ from intergalactic \ion{H}{1} clouds, which
provide upper limits to \phion\ that are 4 times higher than the lower
limit derived here.  However, this model requires a small \fesc\ and a
substantial FUV opacity to attenuate photons capable of ionizing
\ion{Fe}{1} and \ion{Mg}{1}.  Our limit is derived from stringent
3$\sigma$ upper limits to the column density of \ion{Fe}{1} and a
conservative interpretation of the velocity structure observed in Mg I,
Mg II, and Fe II. If this limit is relaxed from 3$\sigma$ to 2$\sigma$
in each undetected \ion{Fe}{1} component then \phion\ $\geq$ 4300
\units.  Similary, if we assume only two undetected components
corresponding to the two seen in the \ion{Mg}{1} line, then
\phion\ $\geq$ 4300 \units.  Thus the assumed limits on \ion{Fe}{1}
produce a very conservative limit on \phion.

If \g\ contributes significantly to the ionizing flux at the absorbing
cloud, then strict constraints on the extragalactic radiation field are
impossible and we are left with the robust limit on total \fesc\ given
by the \ha\ images. The limit \fesc\ $\leq 0.02$ is significantly lower
than previous measurements, but is consistent with theoretical
treatments of escaping ionizing radiation.  However, these two models
predict radically different \fesc, and so should manifest themselves in
distinct fashion. The opacity model proposed by Deharveng et al. (1997)
predicts \fesc\ $\leq 10^{-4} - 10^{-3}$ through a thick, uniform layer
of \ion{H}{1}. Because this model relies exclusively on \ion{H}{1}
opacity, it predicts harder galactic spectra than the geometric model
(Dove \& Shull 1994; Dove et al.~1999). This difference may appear in
the \ion{C}{2} / \ion{C}{4} and \ion{Si}{4} / \ion{C}{4} ratios in
clouds ionized by hardened galactic spectra.  We note that the
currently small sample of galaxies for which limits on \fesc\ have been
obtained is not sufficient to allow these results to be generalized to
all galaxies, or even to all starburst galaxies.  An ensemble average
\fesc\ taken from a larger sample of targets with varying morphology,
viewing geometry, and star formation history is necessary before a
reliable general estimate of \fesc\ can be made. Further steps in this
direction will be taken by the FUSE satellite, which will observe two
low-redshift starbursts shortward of the rest-frame Lyman limit in an
attempt to measure \fesc\ directly.

There are several ways in which future HST observations (with STIS 
or COS) could resolve existing unknowns in this system. 
In addition to estimating the extragalactic spectral index in the LyC,
UV observations of \ion{C}{4} and \ion{Si}{4} with HST/STIS would
determine the overall ionization state of the gas and probe the
existence of a hot, collisionally ionized component.

Due to the beam dilution mentioned above, the measurement $N_{HI} = 8
\times 10^{19}$ cm$^{-2}$ may be an underestimate of the total amount
of \ion{H}{1} present. If so, then this absorber is potentially a
damped Lyman $\alpha$ (DLA) system of very low redshift ($z \sim
0.005$).  High-$z$ DLAs are commonly believed to arise in thick,
extended \ion{H}{1} disks (Wolfe et al.~1995). A key element of this
picture is that the large neutral disks detected as high-$z$ DLAs have
evolved without merging.  This view has been disputed by the
theoretical simulations of Haehnelt et al.~(1998), who argue that DLAs
arise in irregular merging protogalactic clumps. The NGC~3067 cloud, if
it is a DLA, would suggest that these objects can also arise, at least
at low redshift, in interacting systems.  A STIS/FUV observation of the
cloud's Ly$\alpha$ absorption profile along a pencil-beam line of sight
would not be subject to either the 21-cm absorption bias or beam
dilution, and would allow the absorber to be firmly classified.

Finally, a STIS/G140L ($R = 1000$) spectrum could measure many key
abundances (S, Si, C, O, and Ni).  These metal lines should follow a
linear curve of growth, as do the metal lines in the NGC~3067 cloud
presented here.  Based on the numerical models presented here, FUV
lines of \ion{S}{2} at should easily be detectable, as should lines of
\ion{Si}{2}.  The total column density of \ion{S}{2} would constrain
the overall metallicity of the cloud (S and Zn are nearly undepleted in
low-density gas, but Zn II in this cloud is too weak to be detected).
Based on the observed N(Fe II), a high signal-to-noise measurement of
the weak NUV lines of \ion{Fe}{2} could, if unsaturated, improve our
limits on \phion.

Our major results can be summarized as follows: 
\begin{itemize} 

\item For an extragalactic power-law spectrum with $\alpha _{s} ~ = ~
1.8$ we infer \phion\ $\geq$~2600 \units\ and $I_{0}~\geq~10^{-23}$ erg
cm$^{-2}$ s$^{-1}$ Hz$^{-1}$ sr$^{-1}$.

\item Measurements of the ionizing flux from the galaxy NGC 3067 and
\ha\ limits on the ionizing flux at the cloud constrain \fesc\ $\leq$
0.02, independent of $\alpha _{FUV}$ and $\alpha _{s}$.

\item If the incident spectrum is assumed to contain a break at the
Lyman limit, our detection of \ion{Mg}{1} requires that the break be
less than $\sim 150$ for \phion\ $=2600$ \units.

\end{itemize} 

\acknowledgements

This work was based on observations with the NASA/ESA {\it Hubble Space
Telescope} obtained at the Space Telescope Science Institute, which is
operated by AURA, Inc., under NASA contract NAS5-26555.  We thank Gary
Ferland for the use of Cloudy.  This work was supported by {\it HST}
Guest Observer grant GO-05892.01-94A and by the Astrophysical Theory
Program (NASA grant NAGW-766 and NSF grant AST 96-17073).

\pagebreak

\pagebreak

\begin{deluxetable}{cccccc}
\small
\tablenum{1}
\tablewidth{500pt}
\tablecaption{Journal of Observations}
\tablehead{
\colhead{Setting}
&\colhead{Archive}
&\colhead{Wavelength}
&\colhead{Date}
&\colhead{Exposure}
&\colhead{Species}\nl
\colhead{}
&\colhead{Identification}
& \colhead{Coverage (\AA)}
& \colhead{}
& \colhead{Time (s)}
& \colhead{Detected}}
\startdata
 1 & Z2YX0104P   & 2836-2880 & 6/12/96 & 11097.6 & Mg I          \\
 2 & Z2XY0106P   & 2782-2828 & 6/12/96 & 10744.0 & Mg II         \nl
 3 & Z2YX0204M   & 2578-2624 & 4/9/96  & 8976.0  & Fe II         \nl
 4 & Z2YX0206P   & 2340-2388 & 4/9/96  & 12892.8 & Fe II, CIV    \nl
 5 & Z2YX0304T   & 2488-2534 & 3/21/96 & 10472.0 & Fe I (limits) \nl
\enddata
\end{deluxetable}

\pagebreak 
\begin{deluxetable}{lccccccc}
\small
\tablecolumns{8}
\tablenum{2}
\tablewidth{0pt}
\tablecaption{3C 232 Observed Absorption Features\tablenotemark{a}}
\tablehead{
\colhead{}   
&\colhead{}
&\colhead{Rest}
&\colhead{Observed}
&\colhead{Observed}
&\colhead{Equivalent}
&\colhead{Oscillator}
&\colhead{Column} \\
\colhead{Identification}
&\colhead{Setting}
&\colhead{Wavelength}
&\colhead{Wavelength\tablenotemark{e}}
&\colhead{Velocity}
&\colhead{Width}
&\colhead{Strength\tablenotemark{c}}
&\colhead{Density} \\
\colhead{}
&\colhead{}
&\colhead{(\AA)}
&\colhead{(\AA)}
&\colhead{(\kms)}
&\colhead{(m\AA)}
&\colhead{}
&\colhead{(\cd)}
}

\startdata
{Mg I} 2853  &1&2852.96 &2865.96 & 1370/1420 & 280$\pm 220$  & 1.83    &(2.1$\pm 1.0$)\e$^{12}$\nl
{Mg II} 2796 &2&2796.35 &2809.90 & 1300-1600 & 2910$\pm 620$ & 0.6123  &$>$(6.9$\pm 1.5$)\e$^{13}$\nl
             &2&2796.35 &2811.50 & 1617      & 290$\pm 90 $  & 0.6123  &(6.7$\pm 2.3$)\e$^{12}$\nl
{Mg II} 2802 &2&2803.53 &2817.05 & 1300-1600 & 2600$\pm 320$ & 0.3054  &$>$(1.2$\pm 0.2$)\e$^{14}$\nl
             &2&2803.53 &2818.63 & 1620      & 150$\pm 75  $ & 0.3054  &(7.1$\pm 3.6$)\e$^{12}$\nl
{Fe II} 2586 &3&2586.65 &2598    & blend\tablenotemark{d}  & 2070$\pm 210$ & 0.06457 &\nodata\nl
 
{Fe II} 2600 &3&2600.17 &2612.77 & 1453   & 2050$\pm 280$ & 0.2239  &$>$(1.5$\pm 0.2$)\e$^{14}$\nl
{Fe II} 2344 &4&2344.21 &2355.53 & 1448   & 1310$\pm 300$ & 0.1097  &$>$(2.5$\pm 0.8$)\e$^{14}$\nl
{Fe II} 2374 &4&2374.46 &2385.94 & 1450   & 840$\pm 630$  & 0.0313  &(5.4$\pm 4.7$)\e$^{14}$\nl
{Fe I} 2484  &5&2484.02 &\nodata &\nodata & $<$24\tablenotemark{b}  & 0.5569  &$<$8.0\e$^{11}$\nl
{Fe I} 2502  &5&2501.89 &\nodata &\nodata & $<$24\tablenotemark{b}  & 0.04963 &$<$8.8\e$^{12}$\nl
\enddata
\tablenotetext{a}{Tabulated 1$\sigma$ errors contain estimates of error due to continuum placement, 
statistical fluctuations, and uncertainty in the oscillator strength.} 
\tablenotetext{b}{Limiting equivalent widths correspond to 3$\sigma$ features.}
\tablenotetext{c}{Iron oscillator strengths are experimental measurements 
from Bergeson et al.~(1996).
All others are from the compilation by Morton (1991).}
\tablenotetext{d}{\ion{Fe}{2} $\lambda$2586 observed at the velocity of NGC 3067 is blended with 
Galactic \ion{Fe}{2} $\lambda$2600 absorption. } 
\tablenotetext{e}{Observed wavelengths and velocities correspond to the centroid of the 
absorption for unsaturated lines and to a velocity range for saturated lines.} 
\end{deluxetable}


\begin{references}
 
\reference{amd} Armand, C., Milliard, B., \& Deharveng, J.-M. 1994, A\&A, 284, 12 
\reference{feii} Bergeson, S. D., Mullman, K. L., Wickliffe, M. E., Lawler, J. E.,
    Litzen, U., Johansson, S. 1996, \apj, 464, 1044 
\reference{bhm} Bland-Hawthorn, J., \& Maloney, P. R. 1999, \apj, 510, L33  
\reference{bowyer} Bowyer, S. 1991, ARA\&A, 29, 59 
\reference{bruh} Bruhweiler, F. C., Kafatos, M., \& Sofia, U. J. 1986, \apj, 303, L31 
\reference{CvG} Carilli, C. L., \& van Gorkom, J. H. 1992, \apj, 399, 373  
\reference{dea} Deharveng, J.-M., Faisse, S., Milliard, B., \& Le Brun, V. 
          1997 A\&A, 325, 1259  
\reference{isr} De La Pena, M. D., \& Soderblom, D. R. 1997, GHRS Instrument Science
    Report No. 84  
\reference{das} Donahue, M., Aldering, G., \& Stocke, J. T. 1995, \apj, 450, L45  
\reference{ds} Dove, J. B., \& Shull, J. M. 1994, \apj, 423, 196  
\reference{dove} Dove, J. B., Shull, J. M., \& Ferrara, A. 1999, \apj, submitted  
\reference{fardal }Fardal, M. A., Giroux, M. L., \& Shull, J. M. 1998, AJ, 115, 2206 
\reference{hazy} Ferland, G. J. 1996, Hazy, A Brief Introduction To Cloudy, University 
   of Kentucky Department of Physics and Astronomy Internal Report  
\reference{gea} Giallongo, E., Fontana, A., \& Madau, P. 1997, \mnras, 289, 629  
\reference{haardt} Haardt, F., \& Madau, P. 1996, ApJ, 461, 20
\reference{hea} Haehnelt, M.G., Steinmetz, M., \& Rauch, M. 1998, \apj, 495, 647 
\reference{henry} Henry, R. C. 1991, ARA\&A, 29, 89 
\reference{qso} Hewitt, A., \& Burbidge, G. 1993, \apjs, 87, 451  
\reference{hurwitz} Hurwitz, M., Jelinsky, P., Dixon, W. van Dyke 1997, \apj, 481, L31  
\reference{kp-qal} Jannuzi et al. 1998, \apjs, 118, 1  
\reference{kf} Kulkarni, V. P., \& Fall, S. M. 1993, \apj, 413, L63  
\reference{kut} Kutyrev, A. S., \& Reynolds, R. J. 1989, \apj, 344, L9  
\reference{lea} Leitherer, C., Ferguson, H. C., Heckman, T. M., \& Lowenthal, J. D. 1995, 
                 \apj, 454, L19  
\reference{ms} Madau, P., \& Shull, J. M. 1996, \apj, 457, 551
\reference{phil} Maloney, P. 1993, \apj, 414, 41  
\reference{morton} Morton, D. C. 1991, \apjs, 77, 119  
\reference{agn2} Osterbrock, D. E. 1989 {\em Astrophysics of Gaseous Nebulae 
          and Active Galactic Nuclei} (San Francisco: University Science Books) 
\reference{seaton} Seaton, M. J. 1979, \mnras, 187, 73 
\reference{ss} Sembach, K. R., \& Savage, B.L. 1992, \apj, 83, 147   
\reference{sea} Shull, J. M., Roberts, D., Giroux, M. L., Penton, S., \& Fardal, M. 1999, 
             \aj, 118, in press 
\reference{stocke} Stocke, J. T., Case, J., Donahue, M., Shull, J. M., \& Snow, T. P. 
    1991, \apj, 374, 72  
\reference{ss} Sutherland, R. S., \& Shull, J.M. 1999, in preparation 
\reference{vogel} Vogel, S. N., Weymann, R., Rauch, M., \& Hamilton, T. 1995, 
    \apj, 441, 162  
\reference{wfs} Witt, A. N., Friedmann, B. C., \& Sasseen, T. P. 1997, \apj, 481, 809
\reference{wolfe} Wolfe, A.M., Lanzetta, K.M., Foltz, C.B., \& Chaffee, F.H. 1995,
   \apj, 454, 698 
\reference{zheng} Zheng, W., Kriss, G. A., Telfer, R. C., Grimes, J. P., \& 
Davidsen, A. F. 1997, \apj, 475, 469  
\end{references}
\end{document}